\documentclass{book}    
\usepackage{piers}  
\usepackage{amsmath,amssymb,amsfonts,bbm,latexsym,color,bm,mathbbol,mathrsfs}

\newcommand{\scp}[2]{\langle #1 , #2 \rangle}

\newcommand{\ba}{\begin{eqnarray}}
\newcommand{\ea}{\end{eqnarray}}
\newcommand{\bary}{\begin{array}}
\newcommand{\ear}{\end{array}}

\begin{document}

\title{Binary mixtures of chiral gases}
\maketitle

\author {C. Presilla} \author {G. Jona-Lasinio} \affiliation {Sapienza
  Universit\`a di Roma} \address {}
\city {Rome} \postalcode {00185}
\country {Italy} \phone {} 
\fax {} 
\email {} 
\misc {} 
\nomakeauthor

\author {C. Presilla} \author {G. Jona-Lasinio} \affiliation {Istituto
  Nazionale di Fisica Nucleare} \address {}
\city {Rome} \postalcode {00185}
\country {Italy} \phone {} 
\fax {} 
\email {} 
\misc {} 
\nomakeauthor

\begin{authors}

  {\bf C. Presilla}$^{1,2}$ {\bf and G. Jona-Lasinio}$^{1,2}$\\
  \medskip
  $^{1}$Sapienza Universit\`a di Roma, Italy\\
  $^{2}$Istituto Nazionale di Fisica Nucleare, Italy

\end{authors}

\begin{paper}

\begin{piersabstract}
  A possible solution of the well known paradox of chiral molecules is
  based on the idea of spontaneous symmetry breaking. At low pressure
  the molecules are delocalized between the two minima of a given
  molecular potential while at higher pressure they become localized
  in one minimum due to the intermolecular dipole-dipole
  interactions. Evidence for such a phase transition is provided by
  measurements of the inversion spectrum of ammonia and deuterated
  ammonia at different pressures.  In particular, at pressure greater
  than a critical value no inversion line is observed. These data are
  well accounted for by a model previously developed and recently
  extended to mixtures.  In the present paper, we discuss the
  variation of the critical pressure in binary mixtures as a function
  of the fractions of the constituents.
\end{piersabstract}

\psection{Introduction}

According to quantum mechanics chiral molecules, that is, molecules
which are not superimposable on their mirror image, should not exist
as stable stationary states.  Consider ammonia NH$_3$. The two
possible positions of the N atom with respect to the plane of the H
atoms are separated by a potential barrier and can be connected via
tunneling.  This gives rise to stationary wave functions delocalized
over the two minima of the potential and of definite parity. In
particular, the ground state is expected to be even under
parity. Tunneling induces a doublet structure of the energy levels.

On the other hand, the existence of chiral molecules can be
interpreted as a phase transition.  In fact, isolated molecules do not
exist in nature and the effect of the environing molecules must be
taken into account to explain phenomena characterized by
instabilities.  This interpretation underlies a simple mean-field
model developed in \cite{jptprl} to describe the transition of a gas
of $\mathrm{NH}_3$ molecules from a nonpolar phase to a polar one
through a localization phenomenon which gives rise to the appearance
of an electric dipole moment.  Even if ammonia molecules are only
pre-chiral \cite{note1}, the mechanism, as emphasized in
\cite{jptprl}, provides the key to understand the origin of chirality.

A quantitative discussion of the collective effects induced by
coupling a molecule to the environment constituted by the other
molecules of the gas was made in \cite{jonaclaverie}. In this work it
was shown that, due to the instability of tunneling under weak
perturbations, the order of magnitude of the molecular dipole-dipole
interaction may account for localized ground states.  This suggested
that a transition to localized states should happen when the
interaction among the molecules is increased.

Evidence for such a transition was provided by measurements of the
dependence of the doublet frequency under increasing pressure: the
frequency vanishes for a critical pressure $P_\mathrm{cr}$ different
for $\mathrm{NH}_3$ and $\mathrm{ND}_3$. The measurements were taken
at the end of the 1940s and beginning of the 1950s
\cite{Bleaney-Loubster.1948,Bleaney-Loubster.1950,Birnbaum-Maryott.1953b}
but, as far as we know, no quantitative universally accepted
theoretical explanation exists in spite of many attempts.  The model
\cite{jptprl} gives a satisfactory account of the empirical results.
A remarkable feature of the model is that there are no free
parameters. In particular, it describes quantitatively the shift to
zero-frequency of the inversion line of $\mathrm{NH}_3$ and
$\mathrm{ND}_3$ on increasing the pressure.

Recently, we extended the model \cite{jptprl} to gas mixtures
\cite{pjpra}.  This case may be of interest, among other things, for
the interpretation of the astronomical data such as those from Galileo
spacecraft \cite{HSK} which measured the absorption spectrum of
$\mathrm{NH}_3$ in the Jovian atmosphere. Formulas for the critical
pressure of a general mixture have been provided.

In the present paper, we investigate the behavior of the critical
pressure in binary mixtures. We show that the critical pressure of a
chiral species can be increased or decreased by several orders of
magnitude by mixing it with a proper fraction of a proper species,
chiral or non polar.

\psection{Chiral gas}

We model a gas of all equal chiral molecules as a set of two-level
quantum systems, that mimic the inversion degree of freedom of an
isolated molecule, mutually interacting via the dipole-dipole electric
force.  At moderate density, we approximate the behavior of the $N \gg
1$ molecules of the gas with the mean-field Hamiltonian
\begin{align}
  h(\psi)=-\frac{\Delta E}{2}\sigma^x- G \frac{\scp{\psi}{\sigma^z
      \psi}}{N} \sigma^z,
  \label{acca}
\end{align}
where $\sigma^x$ and $\sigma^z$ are the Pauli matrices and $\psi$ is
the Pauli spinor representing the mean-field molecular state with
normalization $\scp{\psi}{\psi}=N$.  The scalar product between two
Pauli spinors is defined in terms of their two components in the
standard way.  The parameter $\Delta E$ is the inversion
energy-splitting measured by spectoscopic methods in the rarefied gas.
The parameter $G$ accounts for the effective dipole interaction energy
of a single molecule with the rest of the gas.  It can be estimated in
two different but equivalent ways \cite{pjpra}.

The first way to estimate $G$ is based on the so called Keesom energy,
namely, $G$ is identified with the effective dipole-dipole interaction
obtained after averaging over all possible molecular distances and all
possible dipole orientations.  These averages are calculated assuming
that, concerning the translational, vibrational, and rotational
degrees of freedom, the $N$ molecules behave as an ideal gas at
thermal equilibrium at temperature $T$. This assumption relies on a
sharp separation (decoupling) between these degrees of freedom and the
inversion motion.  The result is
\begin{align}
  \label{G}
  G = \frac{4\pi}{9}\frac{\mu^4 P}{(4 \pi \varepsilon_0 k_BT)^2 d^3},
\end{align}
where $P$ is the pressure of the gas, $\mu$ the electric-dipole moment
of the molecules and $d$ the minimal distance between two molecules,
namely, the so called molecular collision diameter.  At fixed
temperature, the effective interaction constant $G$ increases linearly
with the gas pressure $P$.

The second way to estimate $G$ is based on the reaction field
mechanism \cite{Bottcher}.  Let us consider a spherical cavity of
radius $a$ in a homogeneous dielectric medium characterized by a
relative dielectric constant $\varepsilon_r$. An electric dipole
$\pmb{\mu}$ placed at the center of the cavity polarizes the
dielectric medium inducing inside the sphere a reaction field
$\pmb{R}$ proportional to $\pmb{\mu}$.  As a result, the dipole
acquires an energy
\begin{align}
  \label{dipole.energy}
  \mathscr{E} = - \frac{1}{2} \pmb{\mu}\cdot\pmb{R} = -
  \frac{\varepsilon_r-1}{2\varepsilon_r+1} \frac{\mu^2}{4 \pi
    \varepsilon_0 a^3}.
\end{align}
Since $\varepsilon_r\simeq 1$, we can approximate the first fraction
in Eq.~(\ref{dipole.energy}) by the Clausius--Mossotti relation
\begin{align}
  \frac{\varepsilon_r-1}{\varepsilon_r+2} = \frac{1}{3} \rho
  \left(\alpha + \alpha^{\mathrm{D}} \right),
\end{align}
where $\alpha$ is the molecular polarizability and
$\alpha^{\mathrm{D}} = \mu^2/(3 \varepsilon_0 k_B T)$ is the Debye
(orientation) polarizability. Observing that for a chiral gas
$\alpha^{\mathrm{D}} \gg \alpha$ (for instance, in the case of NH$_3$
we have $\alpha\simeq 2~\mbox{\AA}^3$ whereas
$\alpha^{\mathrm{D}}\simeq 217~\mbox{\AA}^3$ at $T=300~\mathrm{K}$),
we get
\begin{align}
  \mathscr{E} = - \frac{4\pi}{9} \frac{\mu^4 P}{(4 \pi \varepsilon_0
    k_BT)^2a^3}.
\end{align}
Microscopic arguments \cite{Linder,Hynne-Bullough} show that the
radius of the spherical cavity $a$ is not arbitrary but must be
identified with the the minimum distance between two interacting
molecules, namely, the molecular collision diameter $d$ introduced in
Eq.~(\ref{G}).  We thus have $\mathscr{E}=-G$.

The state $\psi$ collectively describing the inversion degree of
freedom of the gas of $N$ chiral molecules is determined as the
minimal-energy stationary state of the Hamiltionan of
Eq.~(\ref{acca}). This corresponds to find the lowest-energy
eigenstate of the nonlinear eigenvalue problem $h(\psi)\psi = \lambda
\psi$, with the constraint $\scp{\psi}{\psi}=N$.  We refer the reader
to \cite{jptprl,pjpra} for the mathematical details, here we just
state the main results.

There exists a critical value of the interaction strength,
$G_\mathrm{cr}=\Delta E/2$, such that for $G<G_\mathrm{cr}$ the
mean-field eigenstate with minimal energy is $\psi = \sqrt{N}
\varphi_+$, where $\varphi_+$ is the eigenstate of $\sigma^x$ with
eigenvalue $+1$, i.e., the molecules are delocalized.  For
$G>G_\mathrm{cr}$ there are two degenerate solutions of minimal energy
which can be termed chiral states, in the sense they are transformed
into each other by the parity operator $\sigma^x$. For $G \gg
G_\mathrm{cr}$, these solutions become the localized states
$\sqrt{N}\varphi_L , \sqrt{N}\varphi_R$, where $\varphi_L,\varphi_R$
are the eigenstates of $\sigma^z$.  The energy associated with the
state $\psi$ is a continuous function of $G$ with a discontinuous
derivative at $G=G_\mathrm{cr}$.  We thus have a quantum phase
transition between a delocalized (or achiral, or nonpolar) phase and a
localized (or chiral, or polar) phase.  In view of the dependence of
$G$ on $P$, we can define a critical pressure at which the phase
transition takes place
\begin{align}
  P_\mathrm{cr}= \frac{9}{8\pi} \frac{\Delta E d^3 (4 \pi
    \varepsilon_0 k_BT)^2}{\mu^4}.
  \label{pcr}
\end{align}
Note that the value of $P_\mathrm{cr}$ is completely determined in
terms of the microscopic parameters $\Delta E$, $\mu$ and $d$ and the
temperature $T$.

In \cite{jptprl} we have also shown that in the delocalized phase the
inversion angular frequency of the interacting molecules depends on
the pressure as $\hbar \overline{\omega}(P) = \Delta E
\sqrt{1-P/P_\mathrm{cr}}$.  This formula is interesting as it
expresses the ratio of two microscopic quantities,
$\hbar\overline{\omega}$ and $\Delta E$, as a universal function of
the ratio of the macroscopic variables $P$ and
$P_\mathrm{cr}$. Furthermore, it is in very good agreement with some
spectroscopic data showing the shift to zero frequency of the
inversion line of $\mathrm{NH}_3$ or $\mathrm{ND}_3$ at increasing
pressures
\cite{Bleaney-Loubster.1948,Bleaney-Loubster.1950,Birnbaum-Maryott.1953b}.

\psection{Binary mixtures}

Consider a gas mixture of two species labeled 1 and 2.  In this case,
the Clausius-Mossotti relation reads as
\begin{align}
  \frac{\varepsilon_r-1}{\varepsilon_r+2} = \frac{1}{3} \left( \rho_1
    \left(\alpha_1 + \alpha^{\mathrm{D}}_1 \right) + \rho_2
    \left(\alpha_2 + \alpha^{\mathrm{D}}_2 \right) \right),
\end{align}
where the Debye polarization $\alpha_i^{\mathrm{D}} = \mu_i^2/(3
\varepsilon_0 k_B T)$ is given in terms of the molecular
electric-dipole moment $\mu_i$ of the species $i=1,2$.  According to
the reaction field arguments discussed above, a chiral molecule, let
us say of species $i$, having dipole moment $\mu_i$, acquires, due to
the interaction with all the other molecules of the mixture, the
energy
\begin{align}
  \label{Ei}
  \mathscr{E}_i = - \frac{1}{3} \left( \rho_1 \left(\alpha_1 +
      \alpha^{\mathrm{D}}_1 \right) + \rho_2 \left(\alpha_2 +
      \alpha^{\mathrm{D}}_2 \right) \right) \frac{\mu_i^2}{4 \pi
    \varepsilon_0 d_{i}^3} ,
\end{align}
where $d_i$ is the molecular collision diameter of the $i$-th
species. We now specialize the discussion in the following two cases.

\psubsection{Mixtures of a chiral gas with a non polar gas}

For a mixture of chiral and non polar molecules, the mean-field
molecular state of the chiral species, assumed as species 1, is
determined similarly to the case of a single chiral gas.  The only
degree of freedom of the non polar molecules is the deformation which,
in turn, is proportional to the electric dipole moment of the chiral
molecules.  We may thus describe the mixture by a single mean-field
molecular state, $\psi_1$, normalized to the number of molecules of
the species 1, $\scp{\psi_1}{\psi_1}=N_1$.  As before, we assume that
this state is determined as the lowest-energy eigenstate of the
eigenvalue problem associated with the mean-field Hamiltonian
\begin{align}
  h_1(\psi_1)=-\frac{\Delta E_1}{2}\sigma_1^x -G_1
  \frac{\scp{\psi_1}{\sigma_1^z \psi_1}}{N_1} \sigma_1^z.
  \label{accamix}
\end{align}
In this Hamiltonian $-G_1$ represents the effective dipole interaction
energy of a single chiral molecule with all the other molecules,
chiral and non polar, of the mixture. Thus we can identify
$-G_1=\mathscr{E}_1$, where $\mathscr{E}_1$ is given by Eq.~(\ref{Ei})
with $i=1$, $\alpha_1\ll \alpha_1^\mathrm{D}$, and
$\alpha_2^\mathrm{D}=0$, namely,
\begin{align}
  G_1 = \left( \gamma_{11} P_1 + \gamma_{12} P_2 \right),
\end{align}
\begin{align}
  \gamma_{11} = \frac{4\pi}{9} \frac{\mu_1^4}{(4 \pi \varepsilon_0
    k_BT)^2d_{1}^3}, \qquad \gamma_{12} = \frac{1}{3}
  \frac{\alpha_2\mu_1^2}{4\pi\varepsilon_0 k_BT d_{1}^3}.
\end{align}
As usual we used the ideal gas relations $\rho_1=P_1/k_BT$ and
$\rho_2=P_2/k_BT$, where $P_1$ and $P_2$ are the partial pressures of
the two species.

The analysis of the nonlinear eigenvalue problem $h(\psi_1)\psi_1 =
\lambda_1 \psi_1$, with the constraint $\scp{\psi_1}{\psi_1}=N_1$, is
identical to the case of a single chiral gas. We have a localization
phase transition when $G_1=\Delta E_1/2$.  The transition can be
considered as a function of the total pressure $P=P_1+P_2$ of the
mixture and of the fractions of the two species $x_1=P_1/P$ and
$x_2=P_2/P$.  In this case, instead of a unique critical pressure, we
have a critical line parametrized by $x_1$ or $x_2=1-x_1$, e.g.,
\begin{align}
  P_\mathrm{cr} = \frac{\Delta E_1}{2x_1\gamma_{11} +2x_2\gamma_{12}}.
  \label{pcrmix1}
\end{align}
In Fig.~\ref{PcrbinaryNH3mixtures}, we show the variation of the
critical pressure in a NH$_3$--He mixture as a function of the He
fraction.  The value of $P_\mathrm{cr}$ increases from the critical
pressure of pure NH$_3$ to a maximum reached for a vanishing NH$_3$
fraction.  The value of this maximum depends on the nature of the non
polar species, it is greater the smaller is the molecular
polarizability $\alpha_2$.
\begin{figure}[t]
  \includegraphics[width=0.5\columnwidth,clip]{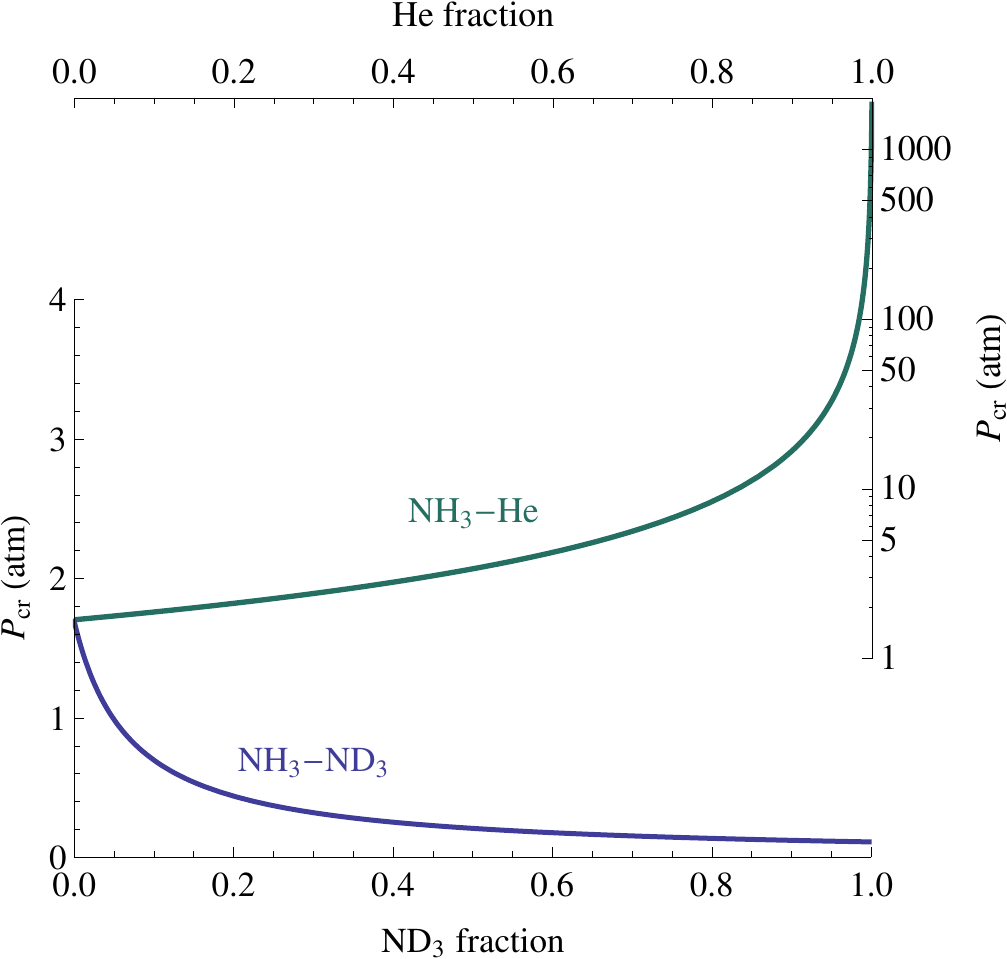}
  \centering
  \caption{Critical pressure of the localization phase transition in a
    binary mixture of NH$_3$ as a function of the fraction of the
    second constituent chosen as ND$_3$ (bottom-left axes) or He
    (top-right axes).}
  \label{PcrbinaryNH3mixtures}
\end{figure}

\psubsection{Mixtures of two chiral gases}

For a mixture of two chiral gases, we describe the inversion degrees
of freedom of the two species by mean-field molecular states
$\psi_1,\psi_2$ normalized to the number of molecules of the
corresponding species.  These states are obtained as the lowest-energy
eigenstates of the eigenvalue problem associated with the coupled
mean-field Hamiltonians
\begin{subequations}
  \label{acca12}
  \begin{align}
    h_1(\psi_1,\psi_2) &= -\frac{\Delta E_1}{2}\sigma_1^x
    -\sum_{j=1}^{2} G_{1j} \frac{\scp{\psi_j}{\sigma_j^z \psi_j}}{N_j}
    \sigma_1^z,
    \\
    h_2(\psi_1,\psi_2) &= -\frac{\Delta E_2}{2}\sigma_2^x
    -\sum_{j=1}^{2} G_{2j} \frac{\scp{\psi_j}{\sigma_j^z \psi_j}}{N_j}
    \sigma_2^z.
  \end{align}
\end{subequations}
Pauli operators now have a label $i=1,2$ relative to the species they
refer to.  Each term $-G_{ij}$ represents the effective dipole
interaction energy of a single molecule of species $i$ with all the
other molecules of species $j$.  By matching $-G_{i1} -G_{i2} =
\mathscr{E}_i$, where $\mathscr{E}_i$ is given by Eq.~(\ref{Ei}) with
$\alpha_1\ll \alpha_1^\mathrm{D}$ and $\alpha_2\ll
\alpha_2^\mathrm{D}$, we get
\begin{align}
  G_{ij} = \gamma_{ij} x_j P, \qquad \gamma_{ij} = \frac{4\pi}{9}
  \frac{\mu_i^2\mu_j^2} {(4\pi\varepsilon_0 k_BT)^2 d_{i}^3},
\end{align}
where $P$ is the total pressure of the mixture and $x_1$, $x_2$ the
fractions of the components.

The solution of the coupled nonlinear eigenvalue problem
$h_1(\psi_1\psi_2)\psi_1 = \lambda_1 \psi_1$ and
$h_2(\psi_1\psi_2)\psi_2 = \lambda_2 \psi_2$, with the constraints
$\scp{\psi_1}{\psi_1}=N_1$ and $\scp{\psi_2}{\psi_2}=N_2$, is
discussed in \cite{pjpra}.  As in the case of a single chiral gas, the
mixture undergoes a localization phase transition at a critical
pressure $P_\mathrm{cr}$.  For $0<P<P_\mathrm{cr}$, the lowest-energy
molecular state of the mixture corresponds to molecules of both
species in a delocalized symmetric configuration.  For
$P>P_\mathrm{cr}$, new minimal-energy molecular states appear with
twofold degeneracy.  These states correspond to molecules of both
species in a chiral configuration of type $L$ or $R$.  We have a
particularly simple formula for $P_\mathrm{cr}$,
\begin{align}
  \label{pcr_chimix_ave}
  \frac{1}{P_\mathrm{cr}} = \sum_{i=1}^{2} x_i
  \frac{1}{P^{(i)}_\mathrm{cr}}, \qquad P^{(i)}_\mathrm{cr} =
  \frac{\Delta E_i}{2\gamma_{ii}} = \frac{9}{8\pi} \frac{\Delta E_i
    d_i^3 (4 \pi \varepsilon_0 k_BT)^2}{\mu_i^4},
\end{align}
the inverse critical pressure of the mixture is the fraction-weighted
average of the inverse critical pressures of its components.  In
Fig.~\ref{PcrbinaryNH3mixtures}, we show the variation of the critical
pressure in a NH$_3$--ND$_3$ mixture as a function of the ND$_3$
fraction.  The value of $P_\mathrm{cr}$ ranges from the critical
pressure of pure NH$_3$ to that of pure ND$_3$, namely, from 1.69 atm
to 0.11 atm.  By changing ND$_3$ with, for instance, D$_2$S$_2$, the
minimal value of $P_\mathrm{cr}$ can be extended down to $4.3 \times
10^{-9}$ atm, the critical pressure of deuterated disulfane.

\psection{Conclusion}

Our approach to the existence of chiral molecules is based on ideas of
equilibrium statistical mechanics. One may be surprised by the
presence of a quantum phase transition at room temperatures. We
emphasize that the transition takes place only in the inversion
degrees of freedom. The dynamics of these degrees of freedom is
affected by temperature only through the values of the coupling
constants.

We have shown that with the addition of a proper fraction of a second
species, non polar or chiral, the critical pressure of an ammonia
mixture can vary in a range of several orders of magnitude.  As a
consequence, the inversion line of ammonia, as well as, possibly, that
of the second chiral constituent, should undergo a frequency shift
rather different from that measured for pure gases, see \cite{pjpra}.
An experimental verification of these predictions is well within the
reach of present technology and would represent a critical test of our
theory.

\end{paper}


\begin{thebibliography}{99}
\bibitem{note1} Ammonia is a potentially chiral molecule, it becomes
  chiral only under substitution of two of the hydrogens with
  different atoms like deuterium and tritium. We refer to potentially
  chiral molecules like ammonia as pre-chiral or, for simplicity,
  chiral.
\end{thebibliography}
\end{document}